\documentclass[prb,floatfix,twocolumn]{revtex4-1} 

\usepackage{amsmath}  % needed for \tfrac, \bmatrix, etc.
\usepackage{amsfonts} % needed for bold Greek, Fraktur, and blackboard bold
\usepackage{graphicx} % needed for figures
\usepackage[capitalise]{cleveref}
\usepackage{tikz}

\begin{document}

\title{Two toy models for the motion of a leaky tank car}
% In a long title you can use \\ to force a line break at a certain location.

\author{Robin Ekman}
\email{robin.ekman@physics.umu.se} % optional
\affiliation{Department of Physics, Umeå University, Ume{\aa } University, SE--901 87 Ume{\aa}, Sweden}

\date{\today}

\begin{abstract}
	I present two toy models for the motion of a tank car as fluid drains out of an off-center opening, based on replacing the fluid with particles.
	No knowledge beyond simple Lagrangian mechanics is required.
	The first toy model is solved analytically and the second can be simulated numerically for a variety of initial conditions and mass ratios.
	Both toy models show the car turning around once as is required by conservation of momentum.
\end{abstract}
\maketitle

\section{Introduction} 

\begin{figure}[b]
	\includegraphics[width=\columnwidth]{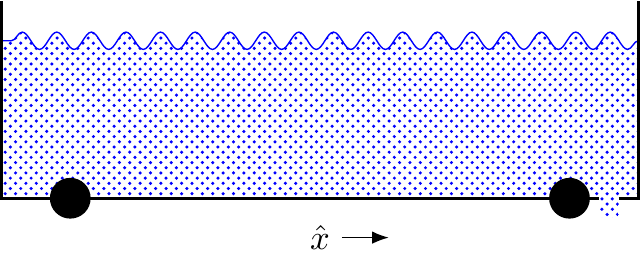}
	\caption{\label{fig:fluid-wagon} A leaky tank car.}
\end{figure}

The leaky tank car problem~\cite{mcdonald1991motion} refers to finding the motion of a tank car filled with liquid, as the liquid drains out of an off-center hole, vertically relative to the car, as in \cref{fig:fluid-wagon}.
Since the center of mass of the liquid moves towards the hole, the tank car must begin to move in the opposite direction to conserve the horizontal momentum of the complete system of tank car and liquid.
Eventually, the tank car must turn around, or all the liquid and the tank car would have acquired momentum in the same direction.

A quantitative description of the motion is precluded by the complexity of fluid flow.
McDonald~\cite{mcdonald1991motion} has analyzed the problem by integrating over the liquid mass, abstracting the fluid flow into just a rate of discharge.
In this paper I present two toy models that replace the liquid with solid particles, reducing the problem to basic Lagrangian mechanics.
%The first toy model is simple enough to be solved analytically, while the second is relatively simple to simulate numerically.
In both cases, the car is found to turn around.

\section{Inclined plane toy model}

\begin{figure}[b]
	\includegraphics[width=\columnwidth]{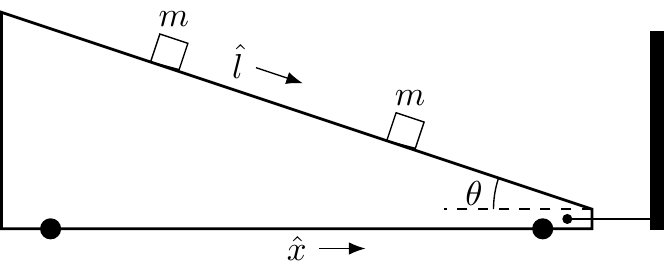}
	\caption{Inclined plane toy model. \label{fig:plane}}
\end{figure}

The first toy model consists of an inclined plane and several blocks that can slide without friction along the plane.
At the low end of the plane is a vertical plate that the blocks will collide with, bringing them to rest relative to the plane, see \cref{fig:plane}

Letting $x$ be the position of the car and $l_i$ be a coordinate along the plane for the $i$:th block, the Lagrangian for this system is \begin{equation}
	L_\text{I} = \frac{M}{2}\dot{x}^2 + \sum_{i=1}^n \frac{m}{2}(\dot{\ell}_i \hat{\ell} + \dot{x} \hat{x})^2 + mg \ell_i \sin \theta .
\end{equation}
Here $\theta$ is the angle of the plane, $M$ is the mass of the car, and $m$ is the mass of each block.
The coordinates $\ell_i$  are taken to increase descending the plane, i.e., the potential energy is \emph{decreasing} with increasing $\ell_i$.

This Lagrangian is valid up to the time $\tau$ when the first block collides with the plate.
At that point, the block is brought to rest relative to the car, i.e,. $\dot{\ell_1} \to 0$.
The impulse received by the block is \begin{equation}
	I = m\left(\dot{x}(\tau_+) - \dot{x}(\tau_- ) - \dot{\ell}_1(\tau_-)  \cos \theta \right)
\end{equation}
where $\tau_\pm$ refers to a quantity just before (after) the collision.
By Newton's third law, the car receives an equal but opposite impulse.
This gives us an equation for $\dot{x}(\tau_+ ) - \dot{x}(\tau_-)$, \begin{multline}
	M\left( \dot{x}(\tau_+) - \dot{x}(\tau_-) \right) \\
 	= -m\left( \dot{x}(\tau_+) - \dot{x}(\tau_- ) - \dot{\ell}_1(\tau_-)\cos\theta \right) \end{multline}
which is readily solved:
\begin{equation}
	\dot{x}(\tau_+) - \dot{x}(\tau_-)  = \frac{m \cos\theta}{M + m} \dot{\ell}_1(\tau_-).
	 \label{eq:impulse}
\end{equation}

After the collision, the system is described by a Lagrangian of the same type, with $n' = n-1$, using $\dot{x}(\tau_+)$ and $\ell_i(\tau), \dot{\ell}_i(\tau)$,  $i>1$, as initial conditions.
\footnote{Because the other blocks do not undergo hard collisions at $t = \tau$, $\ell_i$, $i>0$ is smooth at $t = \tau$ and there's no need to distinguish between $\tau_-$ and $\tau_+$.}

The equations of motion are found as \begin{subequations}
	\begin{align}
		(M + nm) \ddot{x} + m\cos\theta \sum_{i = 1}^n \ddot{\ell}_i & = 0 \\
		m  \ddot{\ell}_i + m \cos\theta \ddot{x}  & = mg \ell_i \sin \theta
		\label{eq:eom}
	\end{align}
\end{subequations}
and can be written in matrix form, 
%\begin{widetext}
\begin{equation}
	\begin{pmatrix}
		M/m + n & \cos \theta &   \cdots &  \cos \theta \\
		 \cos \theta & 1  & \cdots & 0 \\
		% \cos \theta & 0  & \cdots & 0 \\
		\vdots & \vdots  & \ddots & \vdots \\
		 \cos \theta & 0  &  \cdots & 1
	\end{pmatrix}
	\begin{pmatrix}
		\ddot{x} \\
		\ddot{\ell}_1 \\
		%\ddot{\ell}_2 \\
		\vdots\\
		\ddot{\ell}_n
	\end{pmatrix}
	=
	\tilde{g}% \sin \theta
	\begin{pmatrix}
		0 \\
		\ell_1 \\
		%\ell_2 \\
		\vdots \\
		\ell_n 
	\end{pmatrix}
\end{equation}
%\end{widetext}
where $\tilde{g} = g\sin\theta$.
The first equation is a statement of the conservation of momentum in the $\hat{x}$-direction.
This reflects that the Lagrangian is invariant under $x \mapsto x + a$, so that the conjugate momentum is conserved.

The cases $n = 1$ and $n = 2$ are tractable by hand.

\subsection{One block}

The qualitative behavior of the $n = 1$ case, with only one block, is easily understood.
After the block collides with the plate, it is at rest relative to the car.
The only way for momentum to be conserved is if both are also at rest relative to the ground.
That is, as the block slides to the right, the car moves to the left; when the block collides with the plate, the whole system comes to a stop relative to the ground.
Because of Galilei invariance, we realize that if the system is initially moving relative to the ground, the car returns to its initial velocity.

Let us demonstrate this quantitatively.
We solve for $\ddot{x}$ from~\cref{eq:eom} \begin{equation}
	\ddot{x} = - \ddot{\ell} \frac{m}{m + M} \cos \theta
	\label{eq:xddot}
\end{equation}
and substitute into the equation for $\ddot{\ell}$, \begin{equation}
	\ddot{\ell}\left( 1 - \frac{m }{m + M} \cos^2\theta \right) = g \sin \theta 
\end{equation}
Clearly $\dot{\ell} = at + \dot{\ell}(t = 0)$ for a constant acceleration $a$.
From \cref{eq:impulse}, we then have \begin{equation}
	\dot{x}(\tau_+) - \dot{x}(\tau_-) = \frac{m \cos\theta}{M + m}( a\tau + \dot{\ell}(t=0)),
\end{equation}
but~\cref{eq:xddot} gives \begin{equation}
	\dot{x}(\tau_-) = -\frac{m \cos \theta}{M + m}  at + \dot{x}(t = 0)
\end{equation}
and thus \begin{equation}
	\dot{x}(\tau_+) = \dot{x}(t = 0) + \frac{m \cos\theta}{M + m}\dot{\ell}(t=0). \label{eq:1blockconc}
\end{equation}
We have allowed for $\dot{\ell}(t = 0) \neq 0$ as this more general result will be used in the following.

\subsection{Two blocks}

By substituting for $\ddot{x}$ in the equations for $\ell_1, \ell_2$, we find \begin{subequations}
	\begin{align}
		\ddot{\ell}_1 - m \cos^2 \theta \frac{\ddot{\ell}_1 + \ddot{\ell}_2}{M + 2m} & = g \sin \theta \\
		\ddot{\ell}_2 - m \cos^2 \theta \frac{\ddot{\ell}_1 + \ddot{\ell}_2}{M + 2m} & = g \sin \theta
	\end{align}
\end{subequations}
and this system can be solved, e.g. by writing it in matrix form and inverting the $2\times 2$ matrix to find \begin{equation}
	\ddot{\ell}_i = \frac{1}{1 - 2b} g \sin \theta = a_2 
\end{equation}
where $b = m \cos^2 \theta / (M + 2m)$.
The blocks undergo the same acceleration because they are identical and $g$ is constant.
We could have used this from the beginning to derive the same result.

Substituting into the equation for $\ddot{x}$, we find \begin{equation}
	(M + 2m) \ddot{x} + \frac{2 m \cos\theta }{1 - 2b} g \sin\theta = 0.
\end{equation}
and \begin{equation}
	\dot{x} = -\frac{2 m \cos\theta}{M + 2m} a_2
\end{equation}
When the first block hits the plate at $\tau_1$,
\begin{equation}
	\dot{x}(\tau_{1+}) - \dot{x}(\tau_{1-}) = \frac{m \cos\theta}{M + m}
	 a_2 \tau_1,
\end{equation}
so\begin{equation}
\dot{x}(\tau_{1+}) = -\frac{2m \cos \theta}{M + 2m}a_2 \tau_1 + \frac{m \cos \theta}{M + m}  a_2 \tau_1.
\end{equation}

We now use $\dot{x}(\tau_{1+})$ and $\dot{\ell_2} = a\tau_1$ as initial conditions for the one-block case that was treated in the previous section.
This gives us, according to~\cref{eq:1blockconc} \begin{multline}
	\dot{x}(\tau_{2+}) =  \dot{x}(\tau_{1+})  + \frac{m\cos\theta}{M +m} \dot{\ell_2}(\tau_1) = \\
	-\frac{2 m \cos \theta}{M + 2m} a_2 \tau_1  + \frac{m\cos\theta}{M + m} a_2 \tau_1
	+ \frac{m\cos\theta}{M + m} a_2 \tau_1.
\end{multline}
Since $M + 2m > M + m$, this is positive, meaning the car is moving to the right.

However, in this model, the turning is abrupt, as it is the result of the second hard collision against the plate.
In the next section, I will present a model where the motion is smoother.

\section{Curved pipe toy model}

\begin{figure}
	\includegraphics[width=0.33\columnwidth]{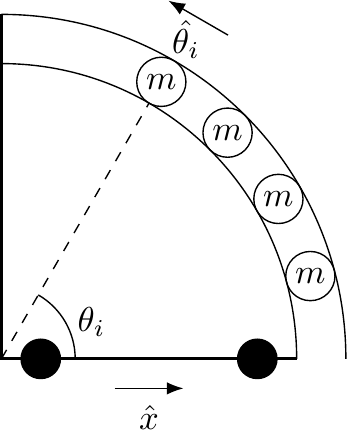}
	\caption{Curved pipe toy model. \label{fig:pipe}}
\end{figure}

This model consists of a curved pipe, specifically in the shape of a quarter circle, mounted to a car.
The pipe bends smoothly to vertical.
Balls of mass $m$ move without friction or rolling in the pipe.
The Lagrangian for this system is \begin{equation}
	L_\text{C} = \frac{M}{2}\dot{x}^2 + \sum_{i=1}^n \frac{m}{2}(r\dot{\theta}_i \hat{\theta} + \dot{x} \hat{x})^2 - mgr \sin \theta_i .
\end{equation}
where $\theta_i$ is the angle from horizontal and $r$ is the radius of the pipe.
$M$ and $m$ are masses, as before.
Clearly $\hat{\theta} \cdot \hat{x} = -\sin\theta$.
The toy model is illustrated in \cref{fig:pipe}.

The equations of motion can be put in matrix form similar to before,
\begin{widetext}
\begin{equation}
	\begin{pmatrix}
		M/m + n &  -r \sin \theta & - r\sin \theta & \cdots & - r\sin \theta \\
		-r \sin \theta & r^2 & 0 & \cdots & 0 \\
		-r \sin \theta & 0 & r^2 & \cdots & 0 \\
		\vdots & \vdots & \vdots & \ddots & \vdots \\
		-r \sin \theta & 0 & 0 &  \cdots & r^2
	\end{pmatrix}
	\begin{pmatrix}
		\ddot{x} \\
		\ddot{\theta}_1 \\
		\ddot{\theta}_2 \\
		\vdots\\
		\ddot{\theta}_n
	\end{pmatrix}
	=
	-gr \begin{pmatrix}
		0 \\
		\cos \theta_1 \\
		\cos \theta_2 \\
		\vdots \\
		\cos \theta_n 
	\end{pmatrix}
	+ \begin{pmatrix}
		r \sum \dot{\theta}_i^2 \cos\theta_i  \\
		\dot{x} \dot{\theta}_1 \cos\theta_1 \\
		\dot{x} \dot{\theta}_2 \cos\theta_2 \\
		\vdots \\
		\dot{x} \dot{\theta}_n \cos\theta_n
	\end{pmatrix}
	\label{eq:c-eom}
\end{equation}
\end{widetext}
but because this system is nonlinear through both products of velocities and  the trigonometric function, it cannot be solved analytically even for $n = 1$.

It is, however, not too difficult to implement a numerical scheme.
Such a scheme solves numerically the equations of motion~\cref{eq:c-eom} until it detects that a ball has reached the bottom of the pipe ($\theta_i < 0$).
It records the time, the positions and velocities at this time, and then continues with one ball fewer, until there are no balls left or the maximum time is reached.

I have written such a code in in Python using the \texttt{numpy} and \texttt{scipy} packages.
The code allows for an arbitrary number of balls and flexible specification of their initial positions.
In the simulations presented here I have set $r = 1$ and $g = 2$, corresponding to measuring lengths in units of the pipe radius and time in units of $t_0$, the free-fall time from a height $r$.
Consequently, velocities are measured in units of $v_0 = gt_0$, half the free-fall velocity.

\begin{figure}[htp]
	\includegraphics[width=\columnwidth]{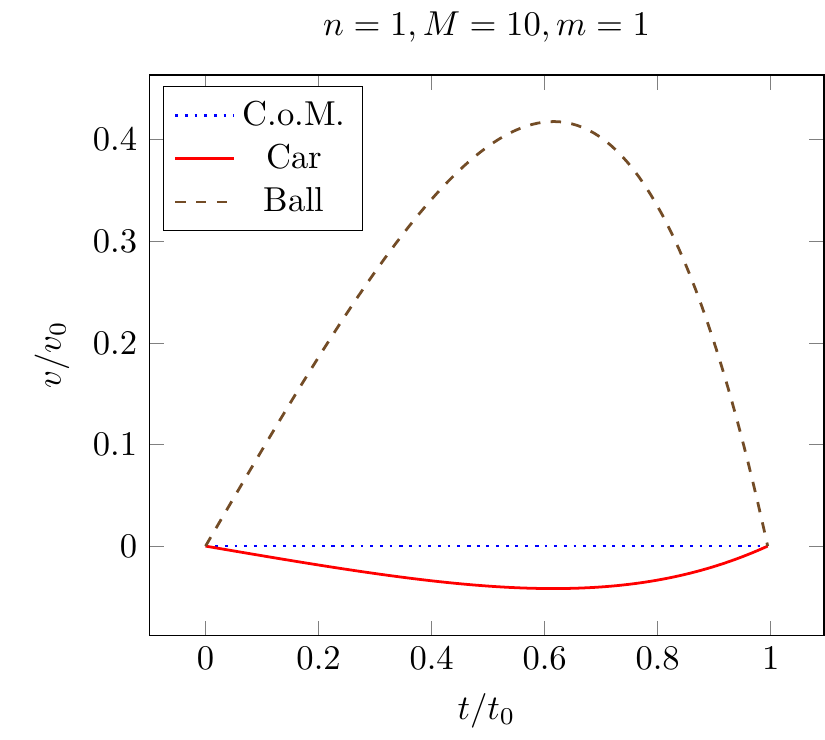}
	\caption{Velocities of the car (solid), ball (dashed) and center of mass (dotted). With a single ball, the car and the ball both end at rest. \label{fig:single}}	
\end{figure}

\begin{figure}[bp]
	\includegraphics[width=\columnwidth]{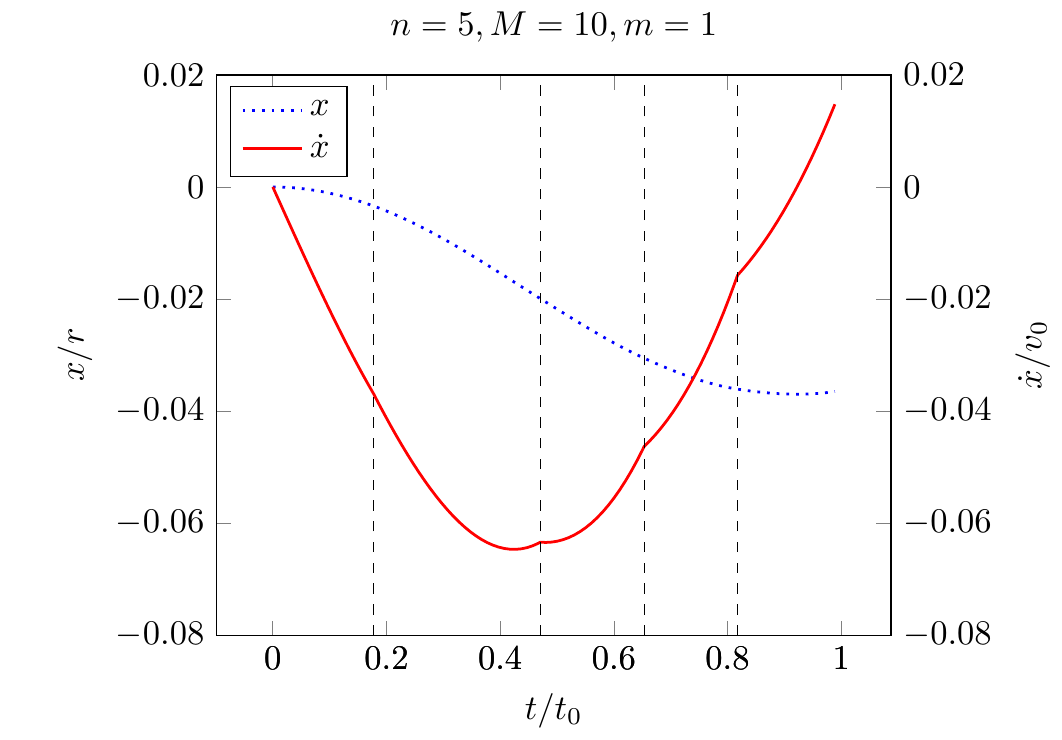}
	\caption{Velocities of the car (solid), one of the balls (dashed) and center of mass (dotted). With multiple balls, the car turns around once.
		Vertical dashed lines indicate when balls fall out of the pipe.
		\label{fig:multiple}}
\end{figure}

\cref{fig:single,fig:multiple} show the output of two simple runs, with a single ball starting at $\theta = \frac{\pi}{4}$ and $n = 5$ balls starting between $\theta = 0.01\pi$ and $\theta = \frac{\pi}{4}$, at equally spaced angles.
We see that, just as in the inclined plane toy model, with a single particle the car returns to rest; with multiple it turns around once.

In \cref{fig:single}, one sees that the center of mass remains at rest to within numerical accuracy.
This is the case for all runs reported here, and is because the first row of \cref{eq:c-eom} is precisely a statement of conservation of horizontal momentum.
 
In \cref{fig:multiple} I have indicated with dashed lines when balls -- other than the last -- fall out of the pipe. 
Looking closely, one can see that there are kinks in the $\dot{x}$ curve at precisely these times, i.e., the acceleration has discontinuities at these times.
While the transition from a circle to a straight line is only once differentiable and one could use a more intricately shaped pipe that transitions smoothly into a straight line instead,\footnote{E.g., let the $x$ and $y$ coordinates be given by a bump function $f$, an infinitely differentiable function with $f(t) = 0$ for $t < a$ and $f(t) = 1$ for $t > b$.} the discontinuities in the acceleration would still remain, because the constraint force from the pipe is discontinuous across $\theta = 0$.

\begin{figure}[bp]
		\includegraphics[width=\columnwidth]{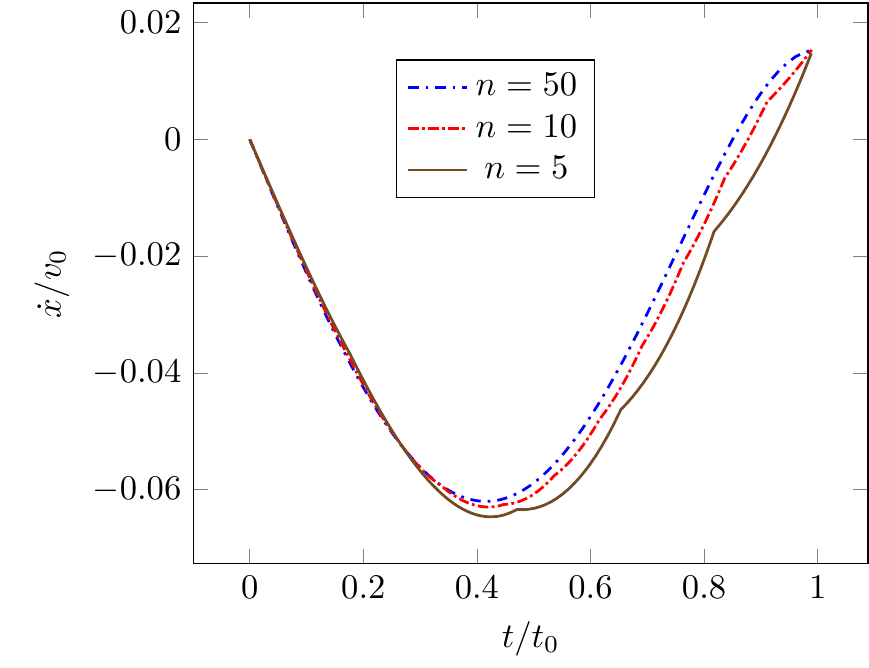}
		\caption{Using a larger number of balls -- with the same total mass --  hides the discontinuous acceleration, but results in similar motion. \label{fig:many}}
\end{figure}

The discontinuities can be made less apparent by instead using a greater number of lighter balls, as in \cref{fig:many}.

\section{Concluding remarks}

The models presented in this paper can be used to illustrate to students the value of highly simplified toy models and a ``try simplest cases'' approach.
In the present case, replacing complex fluid flow with relatively simple particle motion produces toy models that are if not solvable by hand, then at least simple to simulate, and with qualitatively correct behavior.

While the simplest way to obtain the equations of motion is to use the Lagrangian approach, using Newton's second law with constraint forces is also possible, and the models thus require nothing beyond an introductory undergraduate mechanics course.
They could form the basis for a computer lab in such a course.

\begin{acknowledgments}

I would like to thank M. Bradley, G. Brodin, P. Norqvist, and J. Zamanian for introducing me to this problem and fruitful discussions.

\end{acknowledgments}

\nocite{*}
\bibliography{Wagon}{}

\end{document}